\documentclass[prd,nofootinbib]{revtex4}

\usepackage{amsfonts}
\usepackage{amssymb}
\usepackage{bm}

\newcommand{\R}{\mathbb{R}}
\newcommand{\g}{\mathfrak{g}}
\newcommand{\su}{\mathfrak{su}}
\newcommand{\diag}{{\rm diag}}
\newcommand{\csch}{{\rm csch}}

\begin{document}

\title{Explicit $SU(5)$ monopole solutions}
\author{Mark W. Meckes}
\affiliation{Department of Mathematics, Case Western Reserve University,
Cleveland, Ohio 44106.}
\email{mwm2@po.cwru.edu}

\begin{abstract}
We describe in detail a general scheme for embedding several
BPS monopoles into a theory with a larger gauge group, which generalizes
embeddings of $SU(2)$ monopoles discussed by several authors.
This construction is applied to obtain explicit fields for monopoles
with several charge configurations in the
$SU(5)\to [SU(3)\times SU(2)\times U(1)]/\mathbb{Z}_6$ model.
\end{abstract}

\maketitle

\section{Introduction}

Monopoles are topological solitons in three space dimensions in
Yang-Mills-Higgs gauge theories. They arise in theories in which the
gauge group $G$ is spontaneously broken to a residual symmetry group
$H$ by the action of a Higgs field which attains a nonzero vacuum
expectation value (VEV), and in which the vacuum manifold $G/H$ has
non-trivial second homotopy group.
Bogomolnyi-Prasad-Sommerfield monopoles \cite{Bog-76,PS-75}
occur in models with the Higgs field in the adjoint representation, in
the limiting case in which the Higgs potential vanishes, but the
nonzero VEV is kept as a boundary condition. They are static solutions of
a first-order partial
differential equation, the Bogomolnyi equation, and achieve
the lower bound on monopole energy which was derived by Bogomolnyi
\cite{Bog-76}. The simplicity of the Bogomolnyi equation, relative to the
full second-order field equations, makes it easier both to derive
exact solutions, and to study the structure of the moduli space of all
solutions, and a number of methods have been developed to study them (see
\cite{Sut} for a review).

In this paper, we describe in detail how BPS monopoles in several
(possibly different) theories may be embedded simultaneously
into a theory with a larger gauge group. This generalizes a
construction used by a number of authors to embed $SU(2)$ monopoles
into larger theories (see e.g. \cite{Wein}). We apply this embedding
scheme to present explicit fields for several monopoles in the
$SU(5)\to [SU(3)\times SU(2)\times U(1)]/\mathbb{Z}_6$ model.
It is probably well known to experts that an embedding scheme
for BPS monopoles such as is described
here exists, but it has not been described in the literature in this level
of generality; nor has such a scheme been applied to construct monopole
solutions in a phenomenologically interesting model such as the $SU(5)$
model considered here.

The construction here is in a sense complementary to that described by
Lepora in \cite{Lepora}. Lepora discusses embeddings of 't Hooft-Polyakov
monopoles into larger gauge groups away from the BPS limit; by working
in the BPS limit we are able here to present explicit solutions. The
embedding can be treated more simply in the BPS limit as well, by taking
advantage of the fact that the Bogomolnyi equation is almost linear, in a
sense discussed in the next section. Working with the Bogomolnyi equation
allows us to avoid considering topological issues, and also makes trivial
the matter of showing that an embedding does indeed produce a solution.

\section{The general construction}

We consider a model with gauge group $G$ with Lie algebra $\g$.
The Higgs field $\Phi$ is a $\g$-valued scalar field on $\R^3$
and the gauge potential $A$ is a $\g$-valued one-form on $\R^3$.
The field strength is defined by $F=dA+A\wedge A$, and the magnetic
field is $B=\star F$ where $\star$ is the duality operator
on forms. The covariant derivative of the Higgs field is
$D\Phi=d\Phi+[A,\Phi]$. For simplicity of notation, we take the
coupling strength to be unity, and the generators of $G$ to be
antihermitian. Also for simplicity, we work in a fixed gauge.

The Bogomolnyi equation is
\begin{equation}
D\Phi=\pm B. \label{Bog}
\end{equation}
For our purposes, a monopole (or $\g$-monopole) is a pair $(\Phi,A)$
which satisfies the Bogomolnyi equation, along with the boundary conditions
\begin{eqnarray*}
\lim_{r\to \infty}\Phi(0,0,r) &=& \tilde{\Phi} \\
\lim_{r\to \infty}r B\cdot {\bf r} &=& -M
\end{eqnarray*}
for fixed constants $\tilde{\Phi},M \in \g$, where ${\bf r}$ is the
radial vector field on $\R^3$ and $r=\|{\bf r}\|$.
$M$ is called the charge of the monopole, and we call $\tilde{\Phi}$ the
Higgs field at $(0,0,\infty)$. Note that by possibly
making the replacement $\Phi \mapsto -\Phi$, we are free to consider only
the case where the $-$ sign occurs in (\ref{Bog}).

We first wish to note that the Bogomolnyi equation is translation-invariant,
so a monopole may be arbitrarily recentered. Second, note that the
Bogomolnyi equation is invariant under the rescaling
\begin{equation}
(\Phi({\bf r}),A({\bf r}))\mapsto
(\lambda\Phi(\lambda{\bf r}),\lambda A(\lambda{\bf r})) \label{rescale}
\end{equation}
for any constant $\lambda > 0$. This means that we may arbitrarily rescale
the Higgs field at $(0,0,\infty)$. Note also that the rescaling
(\ref{rescale}) leaves the charge $M$ unchanged.

Suppose now that we have $n$ pairwise commuting Lie subalgebras
$\g_1,\g_2,\ldots,\g_n$ of $\g$ (i.e., for every $g\in\g_i$ and
$h\in\g_j$, we have $[g,h]=0$ whenever $i\ne j$) and for each
$i=1,2,\ldots,n$ a $\g_i$-monopole $(\Phi_i,A_i)$ (that is, $(\Phi_i,A_i)$
satisfies the Bogomolnyi equation (\ref{Bog}) for $\g_i$). Let
$\tilde{\Phi}_i$ denote the Higgs field at $(0,0,\infty)$ for $(\Phi_i,A_i)$
and $M_i$ its charge.
(Note that throughout this section, subscripts will always refer to these
subalgebras, not components.) Suppose that we also have a constant
$C\in\g$ which commutes with each of the $\g_i$.
We first make the definitions
\begin{eqnarray}
A&=&\sum_{i=1}^n A_i \\
\Phi&=&C + \sum_{i=1}^n \Phi_i.
\end{eqnarray}
Now since $\g_i$ and $\g_j$ commute for $i\ne j$, $A_i\wedge A_j=0$
and $[A_i, \Phi_j]=0$ in that case. This implies that
$$B=\sum_{i=1}^n B_i\mbox{ and } D\Phi=\sum_{i=1}^n D\Phi_i.$$
Therefore $(\Phi,A)$ satisfies the Bogomolnyi equation (\ref{Bog})
if each of the $(\Phi_i,A_i)$ does.
In this case the Higgs field at $(0,0,\infty)$ is
$$\tilde{\Phi}=C+\sum_{i=1}^n\tilde{\Phi}_i$$
and the monopole has charge
$$M=\sum_{i=1}^n M_i.$$

In more physical terms, what we have is a monopole which is a
superposition of simpler monopoles which live in mutually noninteracting
sectors of the gauge theory. The essential point that makes this work is that
although the Bogomolnyi equation (\ref{Bog}) is nonlinear,
both $B$ and $D\Phi$, and therefore the Bogomolnyi equation,
are additive with respect to sums of fields which commute with each other.

In practice, one begins with the Higgs field at $(0,0,\infty)$, $\tilde{\Phi}$,
and charge $M$ for a monopole and wishes to find corresponding fields
$(\Phi,A)$. To apply the above construction one must first identify
appropriate subalgebras of $\g$ for which monopole solutions are known and
such that their charges add up to $M$.
In order to match the Higgs field at $(0,0,\infty)$, it may be necessary
to rescale the known solutions as in (\ref{rescale}) and add a constant
$C$ which commutes with the subalgebras of $\g$.
We will illustrate this process in the next section. Note that the above
construction does not imply that a monopole corresponding to a particular
Higgs field at $(0,0,\infty)$ and charge must arise from this kind of
embedding.

\section{BPS monopoles in the $SU(5)$ model}

Consider now the symmetry-breaking pattern
$SU(5)\to[SU(3)\times SU(2)\times U(1)]/\mathbb{Z}_6$. This
symmetry-breaking is achieved by a Higgs field with VEV
\begin{equation}
\tilde{\Phi}=i\diag(2,2,2,-3,-3). \label{VEV}
\end{equation}
The fundamental monopole in this model has charge
$$M=\frac{i}{2}\diag(0,0,1,-1,0).$$
To construct the fields for this monopole, we use the following copy of
$\su(2)$ contained in $\su(5)$:
\begin{equation}
\left(\begin{array}{ccccc}
	0&&&&\\&0&&&\\&&\alpha&\beta&\\&&-{\bar \beta}&-\alpha&\\
	&&&&0
	\end{array}\right), \label{1su2}
\end{equation}
and the $\su(2)$-monopole solution found by Prasad and Sommerfield in
\cite{PS-75}. The Prasad-Sommerfield solution is given by
\begin{eqnarray}
\Phi &=& f(r)\sum_{i=1}^3 x_i \tau_i, \label{PS-Phi}\\
A &=& g(r)\sum_{i,j,k=1}^3\epsilon_{ijk}x_i \tau_j dx^k,\label{PS-A}
\end{eqnarray}
where
\begin{eqnarray}
f(r)&=&\frac{1}{r^2}(r\coth r - 1), \label{f(r)}\\
g(r)&=&\frac{1}{r^2}(r\csch r - 1), \label{g(r)}
\end{eqnarray}
and $\tau_j=\frac{i}{2}\sigma_j$. The Prasad-Sommerfield monopole
(\ref{PS-Phi}, \ref{PS-A}) has
$\Phi(0,0,\infty)=\tau_3$ and charge $M=\tau_3$.

Embedding the Prasad-Sommerfield monopole into $\su(5)$ as indicated in
(\ref{1su2}) gives a solution with the desired charge; however to get
the Higgs field at $(0,0,\infty)$ to match (\ref{VEV}), we must rescale
as in (\ref{rescale}) and add some constant matrix $C$ to the Higgs field
which commutes with the copy of $\su(2)$ given by (\ref{1su2}). To see
how to achieve this, we write $C=i\diag(a,b,c,c,-a-b-2c)$, and let
$\lambda$ be the rescaling factor. When we rescale the solution as in
(\ref{rescale}), the Higgs field at $(0,0,\infty)$ gets multiplied by
$\lambda$. Thus to match the Higgs field at $(0,0,\infty)$ we must solve
the equation
$$(2,2,2,-3,-3)=(0,0,\lambda/2,-\lambda/2,0)+(a,b,c,c,-a-b-2c),$$
which has the solution $\lambda=5$, $C=i\diag(2,2,-1/2,-1/2,-3)$.
Rescaling the Prasad-Sommerfield monopole
according to (\ref{rescale}) with $\lambda=5$ and adding $C$ to the
rescaled Higgs field, we obtain the fields
\begin{eqnarray}
\Phi &=& \left(\begin{array}{cccc}
	2i& & & \\
	 &2i& & \\
	 & & -\frac{1}{2}iI_2+\sum_{j=1}^3 25f(5r) x_j \tau_j & \\
	 & & & -3i \end{array}\right) \label{1su2-1}
\\
A &=& \sum_{i,j,k=1}^3 \left(\begin{array}{cccc}
	0& & & \\
	 &0& & \\
	 & & 25g(5r) \epsilon_{ijk} x_i \tau_j & \\
	 & & & 0 \end{array}\right) dx^k, \label{1su2-2}
\end{eqnarray}
where $f(r)$ and $g(r)$ are as given by (\ref{f(r)}) and (\ref{g(r)}),
and $I_2$ is the $2\times 2$ identity matrix.
This solution has been used by Pogosian and Vachaspati in \cite{PV} as a
starting point for numerical study of the interaction of monopoles and
domain walls in the $SU(5)$ model.

We could of course have equally well used another copy of $\su(2)$
for this embedding, such as
\begin{equation}
\left(\begin{array}{ccccc}
	0&&&&\\&\alpha&&&\beta\\&&0&&\\&&&0&\\&-{\bar \beta}&&&-\alpha
	\end{array}\right). \label{1su2.2}
\end{equation}
It is obvious however that this would result in an $\su(5)$ monopole which
is gauge-equivalent to the one constructed above. The same would be true of
any other embedding of the same kind, which embeds $\su(2)$ matrices as
$2\times 2$ submatrices of an $\su(5)$ matrix. To determine whether $\su(2)$
monopoles may be embedded into $\su(5)$ in any other way, we note first
that any such embedding determines a five-dimensional representation of
$\su(2)$. Such a representation must be a direct sum of irreducible
representations. Since the VEV for the $SU(5)$ model has only two distinct
eigenvalues, the same must be true of the Higgs field at $(0,0,\infty)$ for
the embedded $\su(2)$ monopole. But the number of distinct eigenvalues for
any generator for an irreducible representation of $\su(2)$ is equal to the
dimension of the representation.
So the only possible embeddings of an $\su(2)$ monopole into $\su(5)$ must
define a representation of $\su(2)$ which is a direct sum of two-dimensional
representations and trivial representations. Since at most two
two-dimensional representation can fit into $\su(5)$, there are only two
possibilities. If the representation of $\su(2)$ contains only one
irreducible two-dimensional representation, we obtain the embedding discussed
above. If it contains two irreducible two-dimensional representations, then
we obtain a special case of the next embedding to be discussed (namely, the
case in which the two embedded $\su(2)$ monopoles have the same center).

The next monopole of interest in the $SU(5)$ model has charge
$M=\frac{i}{2}{\rm diag}(0,1,1,-1,-1)$. In topological terms, this monopole
has winding two.
In the same way we obtained the fields of the fundamental monopole
above, we could obtain the fields for this monopole by rescaling
and embedding the fields for a monopole in a theory with gauge group
$SU(4)$ and a Higgs field with VEV
$\tilde{\Phi}=\frac{i}{2}\diag(1,1,-1,-1)$, i.e.,
with $SU(4)\to [SU(2)\times SU(2) \times U(1)]/[\mathbb{Z}_2\times
\mathbb{Z}_2]$ symmetry breaking. Unfortunately, the general explicit
solutions for this model are not known.  We can, however, obtain a
family of such solutions by instead using the following two commuting
copies of $\su(2)$ in $\su(5)$:
\begin{equation}
\left(\begin{array}{ccccc}
	0&0&0&0&0\\0&\alpha_1&0&0&\beta_1\\
	0&0&\alpha_2&\beta_2&0\\0&0&-{\bar \beta}_2&-\alpha_2&0\\
	0&\-{\bar \beta}_1&0&0&-\alpha_1
	\end{array}\right), \label{2su2}
\end{equation}
embedding a rescaled Prasad-Sommerfield monopole into each of the positions
indicated by (\ref{2su2}) exactly as was done in (\ref{1su2-1},
\ref{1su2-2}).  In this case, to find rescaling factors $\lambda_1$ and
$\lambda_2$ (one for each embedded $\su(2)$ monopole) and constant
matrix $C=i\diag(-2a-2b,a,b,b,a)$
to make the Higgs field at $(0,0,\infty)$ match, we solve
$$(2,2,2,-3,-3)=(0,\lambda_1/2,0,0,-\lambda_1/2)+
	(0,0,\lambda_2/2,-\lambda_2/2,0)+(-2a-2b,a,b,b,a),$$
and obtain $\lambda_1=\lambda_2=5$, $C=i\diag(2,-1/2,-1/2,-1/2,-1/2)$.
Since each of the $\su(2)$-monopoles may be recentered independently, and
is spherically symmetric,
this construction results in a family of $\su(5)$-monopoles which are
symmetric about the axis connecting the centers of the $\su(2)$-monopoles.
If we assume the $\su(2)$ monopole which is embedded as in (\ref{1su2}) is
centered at the origin and the $\su(2)$ monopole which is embedded as in
(\ref{1su2.2}) is centered at ${\bf c}=(c_1,c_2,c_3)$, then we obtain the
fields
\begin{eqnarray}
\Phi&=&C+\sum_{i=1}^3 25[f(5r)x_i \tau'_i+f(5\|{\bf r}-{\bf c}\|)
	(x_i-c_i)\tau''_i],\\
A&=&\sum_{i,j,k=1}^3 25[g(5r)\epsilon_{ijk}x_i \tau'_j
	+g(5\|{\bf r}-{\bf c}\|)\epsilon_{ijk}(x_i-c_i) \tau''_j]dx^k,
\end{eqnarray}
where $C=i\diag(2,-1/2,-1/2,-1/2,-1/2)$, $\tau'_i$ is $\tau_i$ embedded
in $\su(5)$ as in (\ref{1su2}), and $\tau''_i$ is $\tau_i$ embedded in
$\su(5)$ as in (\ref{1su2.2}).

Finally, we wish to consider monopoles with charge
$\frac{i}{2}\diag(1,1,1,-1,-2)$, i.e., monopoles with winding three.
In this case, we will construct the monopole
by embedding a rescaled Prasad-Sommerfield monopole exactly as in
(\ref{1su2-1},\ref{1su2-2}), and an $\su(3)$ monopole with VEV
$i\diag(1,1,-2)$ and charge $\frac{i}{2}\diag(1,1,-2)$ into the following
$\su(3)$ subalgebra of $\su(5)$:
\begin{equation}
\left(\begin{array}{ccccc}
	\alpha_1 & \beta & 0 & 0 & \gamma \\
	-{\bar \beta} & \alpha_2 & 0 & 0 & \delta \\
	0&0&0&0&0 \\ 0&0&0&0&0 \\
	-{\bar \gamma} & -{\bar \delta} & 0 & 0 & -\alpha_1-\alpha_2
	\end{array}\right),\label{su3}
\end{equation}
(An argument similar to the one outlined above for the fundamental monopole
implies that an embedding of an $\su(3)$ monopole must be of essentially this
type.)
The general solution for such an $\su(3)$ monopole is not known; however
the Higgs field for solutions with spherical symmetry has been found
by Dancer \cite{Dancer}, using the Nahm method \cite{Nahm}. The Nahm data
found by Dancer could in principle also be used to compute the gauge fields
of such monopoles, although this calculation has not been carried out,
but the Higgs field is sufficient to find the energy
density of a monopole, which is proportional to $\Delta{\rm tr}(\Phi^2)$,
where $\Delta$ is the Laplacian on $\R^3$ \cite{Ward}. Dancer's solutions
are given in the form
\begin{equation}
\Phi=i\diag(\Phi_{11}(r),\Phi_{22}(r),\Phi_{33}(r)),
\end{equation}
where
\begin{eqnarray}
\Phi_{11}(r)&=&-\frac{e^{6r}\left(-\frac{3}{2r}+\frac{1}{r^2}-\frac{1}{4r^3}
	\right)+\frac{1}{4r^3}+\frac{1}{2r^2}}{e^{6r}\left(\frac{3}{2r}
	-\frac{1}{4r^2}\right)+\frac{1}{4r^2}},\nonumber\\
\Phi_{22}(r)&=&-(\Phi_{11}(r)+\Phi_{33}(r)),\label{Dancer-Phi}\\
\Phi_{33}(r)&=&\Phi_{11}(-r)\nonumber .
\end{eqnarray}
To match our $\su(5)$ Higgs field at $(0,0,\infty)$, we must rescale Dancer's
solutions by a factor of $\lambda=5/3$ (and also multiply by $-1$ since
Dancer uses the opposite sign convention in the Bogomolnyi equation), and
add to $\Phi$ the constant matrix $C=i\diag(1/3,1/3,-1/2,-1/2,1/3)$.
 We thus obtain the Higgs field
\begin{equation}
\Phi=\left(\begin{array}{cccc}
	-\frac{5}{3}i\Phi_{11}(\frac{5}{3}r)+\frac{1}{3}i & & & \\
	 & -\frac{5}{3}i\Phi_{22}(\frac{5}{3}r)+\frac{1}{3}i & & \\
	 & & -\frac{1}{2}iI_2+\sum_{j=1}^3 25f(5r) x_j \tau_j & \\
	 & & & -\frac{5}{3}i\Phi_{33}(\frac{5}{3}r)+\frac{1}{3}i
	\end{array}\right),
\end{equation}
where $\Phi_{11}$, $\Phi_{22}$, and $\Phi_{33}$ are from (\ref{Dancer-Phi}),
$f(r)$ is given in (\ref{f(r)}), and $I_2$ is again the $2\times 2$ identity
matrix. Again, the $\su(2)$ and $\su(3)$ monopole may be independently
recentered, so that we obtain a family of axially symmetric monopoles.

We have restricted our attention above to embeddings for which explicit
solutions can be given based on solutions in the literature, and which
furthermore are likely to be of interest as a first approximation to
solutions outside the BPS limit (as in \cite{PV} for example). One could
also consider, for example, $\su(5)$ monopoles with charge
$\frac{i}{2}\diag(0,0,k,-k,0)$ which arise by embedding an $\su(2)$
monopole with charge $k\tau_3$ as in (\ref{1su2}), where $k>1$.
It is known that there are such $\su(2)$ monopoles in the BPS limit, but
outside the BPS limit such monopoles are unstable. (In the BPS case, there
is a long-range Higgs attraction between fundamental monopoles which
exactly cancels the magnetic repulsion.) For a review of cases in which
exact solutions are known, see \cite{Sut}.

\begin{acknowledgments} The author would like to thank Prof. T. Vachaspati
for guidance and advice throughout the work that led to this paper, and
for suggesting that this paper be written. The author would also like
to thank Prof. E. J. Benveniste for several helpful discussions.
\end{acknowledgments}

\end{document}